\input harvmac

\def\p{\partial}
\def\ap{\alpha'}
\def\half{{1\over 2}}

\Title{hep-th/0205043}{\vbox{\centerline{Correspondence Principle in a PP-wave
Background}}}
\vskip20pt

\centerline{Miao Li}
\vskip 10pt
\centerline{\it Institute of Theoretical Physics}
\centerline{\it Academia Sinica}
\centerline{\it Beijing 100080} 
\centerline { and}
\centerline{\it Department of Physics}
\centerline{\it National Taiwan University}
\centerline{\it Taipei 106, Taiwan}
\centerline{\tt mli@phys.ntu.edu.tw}

\bigskip

We discuss the correspondence point between a string state and a black hole,
in a pp-wave background, and find that the answer is considerably different
from that in a flat spacetime background. 

\Date{April 2002}

\nref\bmn{D. Berenstein, J. Maldacena and H. Nastase, ``Strings in flat 
space and pp waves from ${\cal N}=4$ Super Yang Mills," hep-th/0202021.}
\nref\ppw{R. Penrose, ``Any spacetime has a plane wave as a limit," in 
Differential Geometry and Relativity, Reidel, Dordrecht, 1976;
J. Figueroa-O'Farrill and G. Papadopoulos, ``Homogeneous fluxes, branes 
and a maximally supersymmetric solution of M-theory," JHEP 0108 (2001)
036,  hep-th/0105308;
M. Blau, J. Figueroa-O'Farrill, C. Hull and G. Papadopoulos, ``Penrose 
limits and maximal supersymmetry," hep-th/0201081.}
\nref\rrm{ R. R. Metsaev, ``Type IIB Green-Schwarz superstring in plane 
wave Ramond-Ramond background," Nucl.Phys. B625 (2002) 70-96,
hep-th/0112044.}
\nref\bhln{V. Balasubramanian, M. Huang, T. S. Levi and A. Naqvi,
``Open Strings from N=4 Super Yang-Mills," hep-th/0204196.}
\nref\suss{L. Susskind, ``Some Speculations about Black Hole Entropy in 
String Theory,"  hep-th/9309145;
E. Halyo, A.Rajaraman and L.Susskind, ``Braneless Black Holes,"
Phys.Lett. B392 (1997) 319-322, hep-th/9605112;
E. Halyo, B. Kol, A. Rajaraman and L. Susskind, ``Counting Schwarzschild 
and Charged Black Holes," Phys.Lett. B401 (1997) 15-20, hep-th/9609075.}
\nref\hp{G. T. Horowitz and J. Polchinski, ``A Correspondence Principle for 
Black Holes and Strings," Phys.Rev. D55 (1997) 6189-6197,  hep-th/9612146;
``Self Gravitating Fundamental Strings,"   Phys.Rev. D57 (1998) 2557-2563,
hep-th/9707170.}
\nref\lms{M. Li, E. Martinec and V. Sahakian, ``Black Holes and the SYM Phase 
Diagram," Phys.Rev. D59 (1999) 044035, hep-th/9809061.}
\nref\mst{J. McGreevy, L. Susskind and N. Toumbas, ``Invasion of the Giant 
Gravitons from Anti-de Sitter Space,"  JHEP 0006 (2000) 008, hep-th/0003075.}

A considerable interest in pp-wave backgrounds has been raised
by the recent work \bmn, which focuses on the correspondence between string theory
in such backgrounds and field theories, which was motivated partially
by earlier works \ppw\ as well as the observation that the string worldsheet theory
is exactly soluble in the light-cone gauge \rrm. It is practically impossible
to cite all relevant references in this letter-type paper, so we refer the 
interested reader to a recent paper \bhln\ for a less than complete list.

We are interested in this work the correspondence point where a string state
turns into a black hole, if the string interaction is present. This scenario
was advocated by Susskind \suss, was made more accurate by Horowitz and 
Polchinski \hp, and was applied to various situations such as matrix theory
\lms. We shall work out the correspondence point in this article, and shall
find that the answer is quite different from the one in a flat background.

A generic pp-wave background of interest to us assumes the form
\eqn\ppmet{ds^2=-4dx^+dx^--z^2(dx^+)^2+dz^2,}
where $x^\pm$ are the light-cone coordinates, $z$ denotes the collection of
transverse coordinates. If the whole spacetime is $D$ dimensional, then
there are $D-2$ components in $z$. Taking $x^+$ as time, its conjugate
is the light-cone Hamiltonian $H=2p^-$, a single string spectrum is given by
\eqn\lgh{2p^-=\sum_n N_{n_i}\left(1+{n_i^2\over (\ap p^+)^2}
\right)^{1/2},}
where each integer $n_i$ corresponds the $n_i$-th Fourier mode of $z_i$,
and the form of the above formula is due to the confining potential
$z_i^2$ for each transverse coordinate.

The typical example is obtained from the Penrose limit of the IIB string
on $AdS_5\times S^5$, when the radius $R$ is taken to infinity.
There are 8 transverse directions. The formula
simplifies considerably in the limit $\ap\rightarrow \infty$ for a fixed
$p^+$, and the light-cone energy in this limit is always an integer. 
On the other hand, for a finite string coupling $g_s$, the effective coupling 
of the super Yang-Mills is infinity, since we are taking the limit 
$N\rightarrow\infty$, so it is impossible to reproduce the result \lgh.
When $N$ is finite, and $R\rightarrow \infty$ as $\ap\rightarrow\infty$,
the coupling on the super Yang-Mills side is finite. Even in this case,
result \lgh\ cannot be reproduced as long as $g_s$ is not vanishing.

So the basic question is, what happens to a generic string state in the
large $\ap$ limit, or in general, to what extent we can trust the perturbative
formula \lgh\ of the string spectrum? There are two possible answers
to this question. One answer is that for a nonvanishing $g_s$, the Newton
gravitational constant becomes large for large $\ap$, so more and more
string states gravitate so much that they become a black hole; another
answer is obvious: for fixed $p^+$, as $\ap\rightarrow \infty$, 
a string state is better described by a giant graviton \mst, since in
the super Yang-Mills language, $J\sim p^+\sqrt{g_sN}\ap \rightarrow
\infty$.  We shall explore the first possibility here, in general for a finite
$\ap$.

In flat spacetime, Susskind proposed sometime ago that when the mass of a string
state gets too large, it will become a black hole \suss, and the degeneracy of
string states roughly matches the degeneracy of black hole states at a given
mass. This proposal was put on a semi-quantitative base by Horowitz and 
Polchinski in \hp, in which they proposed that the turning point between a 
perturbative string state and a black hole is where the horizon size is comparable
to the string scale, this is a place when the worldsheet theory is strongly
coupled. Beyond this point, the black hole is described by a semi-classical
solution to the supergravity equations of motion, and the quantum states can
be no longer accounted for as perturbative string states. This transition
point is called the correspondence point.

Unlike the flat spacetime, the spacetime described by metric \ppmet\ is curved,
with $R_{++}$ nonvanishing. With our convention in \ppmet, coordinate $x^+$
has no dimension, while $x^-$ has the dimension of the length squared, thus
the component $R_{++}$ has no dimension, and is of order 1. We can certainly artificially
introduce a scale $\mu$ by rescaling $x^\pm\rightarrow \mu^{\pm 1}p^\pm$,
so the curvature $R_{++}\sim \mu^2$. But this is just an artificial scale,
for physical quantities, this scale drops out eventually, thus there is
no issue whether the worldsheet theory is strongly coupled or not in
this background. Actually, the action in the lightcone gauge is exactly
soluble. In a sense, the pp-wave background in many ways is similar to
the flat background in which there is no scale.
Thus, we shall be mostly concerned with the extra curvature induced by a string
state, and compare the curvature scale to the string scale.

Due to the confining potential in the transverse directions,
even the massless modes are massive, in the pp-wave background, as can be
seen directly from formula \lgh. When $n_i=0$, some $N_{n_i}=1$, $p^-$
is a constant, so the ``invariant mass" $p^+p^-\sim p^+$ is not
zero in general, thus the force mediated by these massless modes is 
short-ranged, this is drastically different from what happens in a flat
background. In a flat background, the force is long-ranged, thus the
gravitational size of a source is determined directly by the mass and the
Newton constant. Here the story is different, naively we expect that
the potential is Yukawa-like. This is not entirely correct. As we shall
see that in the transverse directions, the force is still Coulomb-like,
but with a different power than in a flat spacetime.

To solve our problem, we may start with the Einstein equations
$G_{\mu\nu}=\kappa^2 T_{\mu\nu}$ where the stress tensor is that of
the string state source. It is almost impossible to solve these equation
exactly, for two reasons. 1. For an excited string, the form of the stress
tensor is complicated. 2. Even for a point-like source, due to the
background \ppmet, we expect the solution be highly asymmetric  between
the longitudinal coordinates and the transverse coordinates.
Thus we will make two simplifications, due to the above two reasons:
1. We will not try to solve the full set of Einstein equations, but
pick out a typical one and replace it by a Poisson-like equation for
a scalar field. 2. We will replace the stress tensor of a string source
by the one of a point-particle source.

We start with a particle action with mass $m$ in a flat background
\eqn\pact{S=\half\int dt\left(e^{-1}(-4\dot{x}^+\dot{x}^-+\dot{z}^2)
-m^2e\right),}
where $e$ is the world-line metric.
Choose the light-cone gauge $x^+=t$, the e.o.m. for $x^-$ implies that
$e$ is a constant, in fact its inverse is proportional to the canonical
conjugate of $x^-$, so we have $p^+=-\half p_-=e^{-1}$. With all this 
in mind and drop the first term which is total derivative, we have
\eqn\plgact{S=\half \int dt (p^+\dot{z}^2-{m^2\over p^+}).}
The last term is of no concern to us, except that the e.o.m. for $e$
induces a constraint
\eqn\constr{4\dot{x}^-=\dot{z}^2+({m\over p^+})^2.}

If the particle is coupled to a generic metric $g_{\mu\nu}$,
we shall replace the first term in \pact\ by
$$\half \int dt e^{-1}g_{\mu\nu}\dot{x}^\mu\dot{x}^\nu,$$
the light-cone gauge fixing can be proceeded in the similar fashion,
if $g_{\mu\nu}$ is a perturbation from the flat metric.
Since the spacetime volume element is $4dx^+dx^-dz=4dtdx^-dz$,
where $dz$ denotes the volume factor of the transverse space, the
energy momentum tensor is simply
\eqn\emten{T^{\mu\nu}={1\over 4}p^+\dot{x}^\mu\dot{x}^\nu\delta(x^--
x^-(t))\delta(z-z(t)),}
where for a free particle, $x^-(t), z(t)$ are linear functions of
$t$. For a particle moving in a pp-wave background, there is a mass
term, or harmonic potential for $z$, this term arises from the term
$z^2(\dot{x}^+)^2$. This term does not contribute to the energy-moment
tensor, but enforces the average value of $\dot{z}$ vanish.
Thus, to the first order approximation for a particle staying in a 
harmonic state we can replace $\delta(z-z(t))$ in \emten\ by
$\delta (z)$.

A string world-sheet action is a little more complicated, in a pp-wave
background
\eqn\sact{S={1\over 4\pi\ap}\int d^2\sigma h^{\alpha\beta}\sqrt{-h}
\left(4\p_\alpha x^+\p_\beta x^- +z^2\p_\alpha x^+\p_\beta x^+
-\p_\alpha z\p_\beta z\right).}
Choose the light-cone gauge $x^+=t$, and $h=-1$. Next, we can follow, say the
steps in Polchinski's book. In the end we have
\eqn\lghac{S=-{l\over \pi\ap}h_{11}\dot{x}^-+{1\over 4\pi\ap}\int d^2\sigma
\left(h_{11}\dot{z}^2-h_{11}^{-1}(z^\prime)^2-h_{11}z^2\right).}
So the conjugate momentum of $x^-$, $p_-$, thus $p^+$ is proportional
to $h_{11}$
\eqn\conp{p^+={l\over 2\pi\ap}h_{11}.}
the factor $h_{11}$ can be absorbed into the definition of $\sigma$, 
so the dependence on $h_{11}$ in the above action completely disappears, 
and the length
of $\sigma$ now is $2\pi\ap p^+$, a dimensionless quantity in our convention.
($x^-$ has the dimension of length squared.) 

The e.o.m. for $h_{\alpha\beta}$ results in
\eqn\eomm{h_{\alpha\beta}\sim 2\p_\alpha x^+\p_\beta x^-
+2\p_\alpha x^-\p_\beta x^++z^2\p_\alpha x^+\p_\beta x^+-
\p_\alpha z\p_\beta z.}
Taking the component $h_{00}$, we find
\eqn\nconst{4\dot{x}^-=\dot{z}^2-z^2+({m\over p^+})^2,}
where the last constant term must be similar to the mass term in \constr,
here $m^2$ is nothing but $p^+p^-$ up to a numerical factor.
Our next strategy is to replace a string by a particle, using the stress
tensor \emten\ with $z(t)=0$, and $x^-(t)$ is determined by \nconst.

We are ready to consider a typical string state. For simplicity, consider the case
when only one transverse coordinate is excited:
\eqn\onet{z=a_n\exp (-i\omega_nt +i{n\sigma\over \ap p^+})+a^\dagger_n
\exp(i\omega_nt -i{n\sigma\over \ap p^+}),}
with
\eqn\freq{\omega_n=(1+{n^2\over (\ap p^+)^2})^{1/2}.}
Suppose the string state correspond to $N_n$ such quanta excited \foot{
A worldsheet constraint is $\sum N_nn=0$, can be met by exciting mode $-n$. 
This won't affect our following discussions.}. 
To estimate the physical size of this string state, we use the
following action
\eqn\modact{S=\int dt p^+(\dot{a_n}^\dagger\dot{a}_n-\omega^2_na_n^\dagger a_n).}
The canonical momenta are $p^+\dot{a}_n$, $p^+\dot{a}_n^\dagger$, so due to Bohr's
quantization condition, we have
\eqn\bohrc{p^+\langle |\dot{a}_n|\rangle \sim {N_n\over \langle |a_n|\rangle}.}
In a bound state, the two terms in the action \onet\ are approximate equal,
namely, $\langle |\dot{a}_n|\rangle\sim \omega_n \langle |a_n|\rangle$.
This together with \bohrc\ gives us the physical size of the string state
\eqn\physize{\langle a_n^\dagger a_n\rangle \sim {N_n\over p^+\omega_n}.}
Of course if we so desire, the above formula can be obtained in a rigorous
way.

We are also ready to estimate the R.H.S. of \nconst. The first term is
about $\omega_n^2\langle a_n^\dagger a_n\rangle$, the second term about $\langle
a_n^\dagger a_n\rangle$. The third term is about $p^-/p^+$, using formula \lgh,
is about $N_n\omega_n/p^+$. This is the same order of the first term, 
according to \physize. In the end, we have
\eqn\lghc{\dot{x}^-\sim {N_n\omega_n\over p^+}.}
This result is somewhat expected, it is equivalent, using the spectrum
\lgh, to saying that $x^-\sim p^-/p^+ x^+$.

We now set out to estimate the gravitational size of the above string state.
Since we are approximating the string by a particle, we need to make sure
that the gravitational size obtained this way should be larger than the physical
size of the string given in \physize. The condition obtained, as we shall see,
is not too strong. The gravitational size is defined as where the perturbation
of the metric becomes of order $1$, if string turns into a black hole, its
horizon is about this size. We will replace one of the Einstein equation
$G_{+-}=\kappa^2T_{+-}$ by the  D'Alembert equation for a scalar
$\nabla^\mu\nabla_\mu\phi =\kappa^2 T_{+-}$, $\phi$ presumably is $h_{+-}$.
The relevant component of the stress tensor is
\eqn\lgstr{\eqalign{T_{+-}&=p^+\dot{x}^+\dot{x}^-\delta (x^--x^-(t))\delta (z)\cr
&=p^+a\delta (x^--at)\delta(z),}}
where 
\eqn\paraa{a\sim N_n\omega_n/p^+.}

The D'Alembert operator in question is
\eqn\lemb{\Delta=-\p_+\p_-+{1\over 4}z^2\p_-^2+\p_z^2.}
Suppose that we know the kernel of this operator
\eqn\kerne{\Delta G(x,y)=\delta^D(x,y),}
then the solution to our equation $\Delta \phi =\kappa^2 T_{+-}$ is given by
\eqn\gens{\phi(x)=ap^+\kappa^2\int dy^+G(x,y^+,ay^+,0),}
where we have identified $y^+$ with time $t$.

Let the Green's function be
\eqn\greenf{G(x,y)=\int {dq_+dq_-\over (2\pi)^2}e^{iq_+(x^+-y^+)+iq_-(x^--y^-)}
G_q(z,w),}
where $z$ is the transverse part of $x$, and $w$ is the transverse part of $y$.
Substituting this into \kerne\ we end up with
\eqn\reker{(\p_z^2-q^{+2}z^2+4q^+q^-)G_q(z,w)=\delta(z,w),}
where we used the relations $q^\pm=-\half q_\mp$.
Let $\phi_n(z)$ be normalized eigen-functions of the operator in the L.H.S. 
of the above equation, which happen to be real, the delta function can be written
as 
\eqn\delt{\delta (z,w)=\sum_n \phi_n(z)\phi_n(w).}
Let 
$$G_q(z,w)=\sum_n a_n(w)\phi_n(z),$$
and substitute the above and \delt\ into \reker, we find
\eqn\findr{\sum_n \left(4q^+q^--|q^+|\sum_i(2n_i+1)\right)a_n(w)\phi_n(z)
=\sum_n\phi_n(w)\phi_n(z),}
this determines $a_n(w)$ and thus
\eqn\tgreen{G_q(z,w)=\sum_n {\phi_n(w)\phi_n(z)\over 4q^+q^--|q^+|\sum_i(2n_i+1)-i
\epsilon}.}
Note that $n$ denotes a set of integers $n_i$ each corresponding to the $n_i$-th
eigen-value of the harmonic oscillator in the $z_i$ direction. We also
introduced the $i\epsilon$ prescription to take causality into account.

Using the Green's function obtained above in \gens\ and integrating over
$y^+$
\eqn\comr{\phi(x)=-{ap^+\over\pi}\kappa^2\int dq^+e^{2iq^+(ax^+-x^-)}\sum_n
{\phi_n(0)\phi_n(z)\over 4aq^{+2}+|q^+|\sum(2n_i+1)+i\epsilon},}
where $\phi_n(0)$ is a constant depending only on $q^+$. The above integral seems to 
have a pole at $q^+$, actually this pole is removed by factors in functions
in the numerator of the fraction in \comr. A normalized eigen-function
is given by
\eqn\hertm{\phi_n(z)=({|q^+|\over\pi})^{{D-2\over 4}}\prod_i {1\over \sqrt{n_i!}2^{n_i/2}}
H_{n_i}(\sqrt{|q^+|}z_i)e^{-\half |q^+|z_i^2},}
$H_{n_i}$ are Hermite polynomials. We see that there is a factor
\eqn\nume{({|q^+|\over \pi})^{{D-2\over 2}}}
in the integral of \comr, as long as $D\ge 4$, there will be no pole at $p^+=0$
in this integral. Also, if $n$ is an odd integer, $H_n(0)=0$, so only
even integers $n_i$ appear in \comr.
It appears impossible to perform the integral in
\comr, moreover, it is hard to sum up the series in $n$, therefore
we shall be content with a rough estimate.

Let us concentrate on the region where $x^--ax^+=0$, close to the trajectory of the string
in the light-cone directions. Taking, for example, $D=10$, the first integral
in the sum of \comr\ reads
\eqn\finte{I=\int_0^\infty dq {q^3\over 4aq+8}e^{-\half q z^2},}
where $z^2=\sum z_i^2$. 
To estimate this integral, we use the steepest-descent method. For fixed $z$, the function
\eqn\sfun{ f(q)=\half z^2q-3\ln q +\ln (aq+2),}
attains the minimal value at
\eqn\miniv{q_0={1\over z^2}\left(2-{z^2\over a}+[({z^2\over a})^2+8{z^2\over a}+4]^{1/2}
\right).}
Up to the second order, the integral \finte\ is approximated by
\eqn\wkbs{I={1\over 4}\sqrt{{2\pi\over f''(q_0)}}e^{-f(q_0)}.}
The first term in the sum of \comr\ is about $\phi_1(z)=ap^+\kappa^2 I$. 
Consider two extreme limits. In one limit, $z^2\gg a$, $q_0$ is
approximately $6/z^2$, and $f''(q_0)$ is approximately $z^4/12$,
so
\eqn\fiphi{\phi_1={ap^+\kappa^2\over z^8}}
up to a numerical coefficient. This is in contrast to the Coulomb potential 
$1/z^7$ in 9 spatial dimensions.
Next consider the other extreme limit $z^2\ll a$. In this case, $q_0$ is about
$4/z^2$, and $f''(q_0)$ is about $z^4/8$. In this limit, we have, up to
a numerical coefficient
\eqn\sephi{\phi_1={p^+\kappa^2\over z^6}.}
This is more like a Coulomb potential in 8 spatial dimensions.

To appreciate the physical meaning of the above results, let us substitute the
approximate result $a\sim N_n\omega_n/p^+$ into relevant formulas. 
In case when $z^2\gg a$, we have the gravitational size
\eqn\gsize{z^2\sim (N_n\omega_n)^{1/4}l_p^2,}
where we replaced $\kappa^2$ by $l_p^8$, $l_p$ is the Planck length.
To be consistent, this transverse gravitational size must be much greater
than $a$, namely
\eqn\pcond{p^+\gg (N_n\omega_n)^{3/4}l_p^{-2}.}
Apparently, this is valid in the large $p^+$ limit. In particular, when
$\ap p^+\gg 1$, $\omega_n\sim 1$, and the above condition is a further constraint
on the largeness of $p^+$. Also, when $z^2\gg a$, the transverse gravitational
size is much greater than the physical size given in \physize, and our assumption
that the string state can be replaced by a particle is justified.
The Horowitz-Polchinski correspondence point is where when $z^2\sim \ap$. 
Using \gsize\ we find
\eqn\corrs{g_s^2\sim {1\over N_n\omega_n},}
to be contrasted to the flat spacetime result $g_s^2\sim 1/\sqrt{N_nn}$ \hp.
To summarize, we have found that when $p^+$ is sufficient large, as physically
reasonable condition, the string state turns into a black hole when condition
\corrs\ is satisfied, or when the condition is turned into an inequality with 
the L.H.S. greater. The giant graviton condition, in terms of $p^+$,
is \bmn
\eqn\gicond{p^+\ge g_s^{-1/2}l_p^{-2}.}
At the correspondence point, it is just
\eqn\gic{p^+\ge (N_n\omega_n)^{1/4}l_p^{-2}.}
It is certainly satisfied if \pcond\ is satisfied.
We see that the string state first turns into a giant graviton, then later when
condition \pcond\ is met, turns into a black hole.

It is straightforward to generalize the above discussion to any other dimension
$D$.

One may start to worry whether our estimate is valid,
since a string state first turns into a giant
graviton, and later turns into a black hole when we dial the coupling constant.
Note that the only place we have used the string property is in the estimate of
the constant $a$, which in general for any kind of state may be replaced by
$p^-/p^+$.  Thus, we need only to replace $N_n\omega_n$ in all formulas above
by $p^-$, we are in a save position.

Next we consider the other limit $z^2\ll a$. The transverse gravitational size
is
\eqn\sgsize{z^2\sim (p^+\kappa^2)^{1/3}.}
In order to be consistent, 
\eqn\ccond{p^+\ll (N_n\omega_n)^{3/4}l_p^{-2},}
a small $p^+$ limit, just opposite to \pcond. Also, for our point-particle approximation
to be valid, $z^2$ must be larger than the physical size, thus the condition
\eqn\acond{p^+\ge ({N\over \omega_n})^{3/4}l_p^{-2},}
this is consistent with \ccond\ if $\omega_n$ is sufficiently large, and is the case
for small $p^+$, see eq.\freq. The correspondence point $z^2\sim \ap$ occurs
at 
\eqn\occc{p^+\sim {1\over g_s^2\ap}.}
Compared to the giant graviton point \gicond, or $p^+\sim 1/(g_s\ap)$, again
for small $g_s$, the string state turns into giant graviton first, and later
becomes a black hole.
Interestingly, at the correspondence point, when we apply the condition \ccond\
to \occc, we find
\eqn\acorrs{g_s^2\gg {1\over N_n\omega_n},}
quite similar to the condition \corrs.
It is possible, then, that the correspondence condition \corrs\ is quite general
and is not limited to extremely large $p^+$ or extremely small $p^+$.

So far we have concentrated on the first term in \comr\ at $x^--ax^+=0$. It is
harder to estimate other terms, even at $x^--ax^+$. It is reasonable to assume
that these terms are subleading when $z^2$ is large. In a WKB approximation,
the dominant $q^+$ would behave as $1/z^2$, and the leading Coulomb potential
$1/z^8$ will not be modified, as one can easily see in \comr.

Away from $x^--ax^+=0$, the exponential $e^{2iq^+(ax^+-x^-)}$ plays the role of
a damping factor, thus the dominant $z$ will be smaller. Note that $x^--ax^+=0$
is the longitudinal center of the gravitational field generated by the string
state. Viewing the horizon as a deformed sphere, then what we have inspected
is the transverse size of the horizon along the equator of this sphere. The
south pole and the north pole correspond to points where $z=0$. When the
damping factor is included, it is not possible even to determine $q_0$ in
the first term in \comr\ in the WKB method, since the function $f(q)$ now
includes a new term $-\ln\cos(2qx^-)$ (setting time $x^+=0$). We are thus
content with a rough estimate. For large $z^2>|x^-|$, apparently we still have
$q_0\sim 1/z^2$, the picture we have had so far is not much modified. 
For small $z^2$, namely when $z^2\ll |x^-|$, and the gravitational field
behaves as
\eqn\lga{\phi_1\sim {ap^+\kappa^2\over (x^-)^4},}
this is certainly expected. Again, the gravitational size at these poles
are the same order as the one along the equator, since $x^-$ has the dimension
of the length squared. 

Finally, the degeneracy of states at the correspondence point, or the entropy.
For large $z^2\gg a$, we argued above that both the transverse size and the
longitudinal size are the same order of magnitude, so the entropy is, according
to Bekenstein-Hawking formula, $S\sim z^8/l_p^8$. At the correspondence point,
$z^2\sim\ap$, so $S\sim 1/g_s^2$, and by virtue of \corrs, we have
\eqn\entr{S\sim N_n\omega_n=p^-,}
this is greater than the entropy in the flat background at the same $N_n$.
It would be interesting to work out this result on the super Yang-Mills side.

Acknowledgments. 
This work was initiated at the Isaac Newton Institute at Cambridge, England,
when the author participated the M-theory workshop, the organizers
of that workshop and the staff of the Newton Institute are gratefully
acknowledged.
This work was supported by a grant of NSC, and by a 
``Hundred People Project'' grant of Academia Sinica and an outstanding
young investigator award of NSF of China.

\vfill
\eject

\listrefs
\end